\documentclass[preprint,5p,twocolumn]{elsarticle}

\usepackage{amssymb}
\usepackage{amsmath}
\usepackage{aas_macros}
\usepackage{hyperref}

\journal{Journal of High Energy Astrophysics}

\begin{document}

\begin{frontmatter}



\title{Constraining the parameters of heavy dark matter and memory-burdened primordial black holes with DAMPE electron measurements.}

\author[1]{Tian-Ci Liu} 
\author[2]{Ben-Yang Zhu}
\author[1]{Yun-Feng Liang\corref{cor1}}
\author[3,4]{Xiao-Song Hu}
\author[1]{En-Wei Liang\corref{cor2}}

\cortext[cor1]{liangyf@gxu.edu.cn}
\cortext[cor2]{lew@gxu.edu.cn}

\affiliation[1]{organization={Guangxi Key Laboratory for Relativistic Astrophysics, School of Physical Science and Technology},
            addressline={Guangxi University}, 
            city={Nanning},
            postcode={530004}, 
            country={China}}
\affiliation[2]{organization={Key Laboratory of Dark Matter and Space Astronomy},
            addressline={Purple Mountain Observatory, Chinese Academy of Sciences}, 
            city={Nanjing},
            postcode={210023}, 
            country={China}} 
\affiliation[3]{organization={School of Physics and Astronomy},
            addressline={Beijing Normal University}, 
            city={Beijing},
            postcode={100875}, 
            country={China}}
\affiliation[4]{organization={Faculty of Arts and Sciences},
            addressline={Beijing Normal University}, 
            city={Zhuhai},
            postcode={519087}, 
            country={China}}

\begin{abstract}
The DArk Matter Particle Explorer (DAMPE) is a space-based instrument for detecting GeV-TeV cosmic rays and gamma rays. High-energy cosmic rays could be emitted from several dark matter candidates theoretically, such as the heavy dark matter (HDM) and the primordial black holes (PBHs). 
HDM particles with a mass of $>100\,{\rm TeV}$ could decay into $\gtrsim 10$ TeV electron/positron pairs. PBHs with a mass of $\lesssim 10^{10}\,{\rm g}$ would survive to the present day if the Hawking radiation is significantly suppressed due to the memory burden effect and can also lead to the emission of $\gtrsim 10$ TeV electrons. 
In this work, we use the DAMPE electron measurements to obtain the constraints on the decay lifetime $\tau$ of HDM and the entropy index $k$ of memory-burdened PBHs
at $95 \%$ confidence level. The constraints on the fraction $f_{\rm PBH}$ are also derived with a fixed $k$. 
Furthermore, the high-energy tail of the DAMPE electron spectrum shows a sign of going upwards, possibly suggesting the presence of an additional component; we discuss if this spectral behavior is real, which parameter space is required for it to be attributed to HDM or PBH.
We will show that the required parameters have been constrained by existing limits.
\end{abstract}




\end{frontmatter}



\section{Introduction}
\label{sec1}

Dark matter (DM) has been studied for decades and the existence of which has been supported by many evidences \cite{1991MNRAS.249..523B,2006ApJ...648L.109C,2016A&A...594A..13P}. However, the nature of DM is still unknown. A lot of DM candidates involving new physics have been proposed to explain the nature of DM, such as weakly interacting massive particles (WIMPs) \cite{1996PhR...267..195J,2010ARA&A..48..495F,2018PhRvL.121k1302A}, axion-like particles (ALPs) \cite{2016PhR...643....1M,2018PrPNP.102...89I,2000PhRvL..85.1158H}, and primordial black holes (PBHs) \cite{1975CMaPh..43..199H,2014MPLA...2940005B,2024PhRvD.110f3021L}.

Heavy dark matter (HDM) is also a kind of DM candidate proposed in several models, including the supergravity model \cite{2008PhRvL.100f1301I}, WIMPzillas \cite{1999AIPC..484...91K}, gravitinos \cite{1982PhRvL..48..223P}, glueballs \cite{2017PhRvD..95d3527H} and so on. HDM could annihilate or decay into the Standard Model (SM) particles where the spectra depend on the DM particle mass and channels. Based on these annihilation or decay signals, one can search for and constrain HDM indirectly, which has been widely studied in many researches \cite{2022PhRvL.129z1103C,2016PhRvD..94f3535K,2024PhRvD.109f3036H,2020JCAP...01..003I}.

Primordial black hole (PBH) is a hypothetical object formed in the early universe, where the energy density is so dense that the matter could collapse to a black hole with a $\delta \sim 1$ fluctuation \cite{2021RPPh...84k6902C,2020ARNPS..70..355C}. Because of the special forming process, PBHs have an extremely wide mass range from Planck mass to $\sim 10^5$ $M_{\odot}$ depending on the formation time. 
It has long been believed that, since the evaporation rate of the Hawking radiation is inversely proportional to the PBH mass squared, $M\lesssim 5\times10^{14}$ g PBHs would be totally evaporated and could not survive today as a DM candidate. Recently, \citet{2020PhRvD.102j3523D} proposed that the Hawking radiation of small mass PBHs would be significantly suppressed due to the memory burden effect. This effect opened a new mass window that DM could be totally composed of PBHs in a mass range of $\lesssim10^{10}$ g \cite{2024MNRAS.532..451T}.

The high-energy electrons emitted from above DM candidates (HDM and PBHs) could be detected by the Fermi-LAT telescope \cite{2012PhRvL.108a1103A}, the Alpha Magnetic Spectrometer (AMS-02) \cite{2013PhRvL.110n1102A}, the CALorimetric Electron Telescope (CALET) \cite{2018PhRvL.120z1102A}, the Large High Altitude Air Shower Observatory (LHAASO) \cite{2019arXiv190502773C}, the High Energy Stereoscopic System (HESS) \cite{2024PhRvL.133v1001A}, the IceCube neutrino observatory \cite{2019PhRvD.100h2002A}, and the DArk Matter Particle Explorer (DAMPE) \cite{2017Natur.552...63D}.
DAMPE is a space-based high-energy particle detector for studying cosmic rays and gamma rays up to $\sim 10$ TeV with good energy resolution and large acceptance, which is suitable for detecting indirect DM signals \cite{2017APh....95....6C,2017Natur.552...63D,2022SciBu..67..679D,2023PhRvD.108f3015C,2022SCPMA..6569512L,2024JHEAp..44..210M,2024JHEP...10..094Y}.
In this work, we use the DAMPE electron measurements in the energy range of 55 GeV to 4.57 TeV \cite{2017Natur.552...63D} to constrain the parameters of HDM and memory-burdened PBHs. The decay lifetime of HDM is constrained to be longer than $\sim 10^{27}$ s in the mass range of $10-10^{5}$ TeV at $95 \%$ confidence level.
The entropy index $k$ in the suppressed factor of the Hawking radiation from memory-burdened PBHs is constrained to be close to the most stringent result in the mass range of $10^{3}-10^{10}$ g at $95 \%$ confidence level.  
Constraints on the fraction $f_{\rm PBH}$ are also derived with a fixed $k$ as an example.

This paper is structured as follows. The HDM and memory-burdened PBHs are briefly introduced in Sec. \ref{sec2}. In Sec. \ref{sec3}, the constraining method is described. We present the final results in Sec. \ref{sec4} and conclude in Sec. \ref{sec5}.

\section{Dark matter}
\label{sec2}

\subsection{Heavy dark matter}

HDM has a mass ranging from 1 TeV to even the Planck mass ($\sim 10^{16}$ TeV), indicating that high-energy particles from HDM decay would fall within the DAMPE sensitivity range. Although HDM has been discussed in many models \cite{2008PhRvL.100f1301I,1999AIPC..484...91K,1982PhRvL..48..223P,2017PhRvD..95d3527H}, only the decay nature is considered in this work phenomenologically as a model-independent situation.

HDM could decay into electrons directly ($\rm e^{+}e^{-}$ channel),
\begin{equation}
    \frac{{\rm d}N_{\rm e}}{{\rm d}E}=\delta (E-\frac{m_{\chi}}{2}),
\end{equation}
or decay into various SM particles like quarks and leptons, and then further decay or hadronize into electrons (other channels) \cite{2005PhR...405..279B}.

\subsection{Memory-burdened PBHs}

PBHs are a kind of massive compact halo objects that formed from the spherical collapse when the fluctuations reenter the particle horizon at the early universe \cite{2022JHEP...08..001M}. 
The PBH mass is proportional to the forming time after the Planck time and before the Big Bang nucleosynthesis (BBN). That means the PBH mass could be extremely low (even down to the Planck mass $\sim 10^{-5}\,\rm g$) and the Hawking radiation would be effective.
However, PBHs as the DM candidate must be $M\gtrsim 5\times10^{14}$ g so that they could survive to date from the Hawking evaporation \cite{2010PhRvD..81j4019C}.

\citet{2020PhRvD.102j3523D} proposed the memory burden effect on PBHs, indicating that the status of a half-decayed PBH is different from a new one of the equal mass. 
With the PBH evaporating, the surface area would shrink while the quantum information is conserved in black holes since the Hawking radiation is thermal. Therefore, the decreasing entropy would give feedback such that the electron spectra from evaporation are significantly suppressed by a factor of $S^{k}$ \cite{2024MNRAS.532..451T},
\begin{equation}
    \frac{{\rm d}^{2} N_{\rm e}}{{\rm d}E{\rm d}t}=\frac{1}{S^{k}} \frac{\Gamma_{\rm e}}{{2\pi\hbar}[\exp\left({{E}/{k_{\rm B}T}}\right)+1]},
\end{equation}
with the Bekenstein-Hawking entropy in units of the Boltzmann constant
\begin{equation}
    S=\frac{4\pi GM^{2}}{\hbar c} \approx 2.6 \times 10^{10} \left(\frac{M}{1\,\rm g}\right)^2,
\end{equation}
where $M$ and $k$ are the PBH mass and the entropy index. The second part is the primary emission of traditional Hawking evaporation, with the graybody factor $\Gamma_e$ and the temperature of black holes $T$ \cite{2010PhRvD..81j4019C,1990PhRvD..41.3052M}. The emitted particles would also decay or hadronize into secondary electron emission. 

The memory burden effect on PBHs opened a new mass window in the mass range of $\lesssim10^{10}$ g. \citet{2024MNRAS.532..451T} derived the new constraints from Galactic gamma-ray emission, Extragalactic gamma-ray background, BBN, and cosmic microwave background (CMB) anisotropies. The constraining result from high-energy neutrinos is also derived in \cite{2024arXiv241007604C}.

\section{Method}
\label{sec3}

The number density of electrons emitted from HDM decay or PBH Hawking radiation could be derived from the propagation equation \cite{2006A&A...455...21C}, 
\begin{equation}
D(E)\nabla^2\left(\frac{{\rm d}n_{\rm e}}{{\rm d}E}\right)+\frac{\partial}{\partial E}\left[b(E)\frac{{\rm d}n_{\rm e}}{{\rm d}E}\right]+Q_{\rm e}=0,
\label{eq:prop}
\end{equation}
where $D(E)$ is the diffusion coefficient, $b(E)$ is the energy-loss term, and $Q_{\rm e}$ is the source injection. 

For HDM, $Q_{\rm e}$ is expressed as \cite{2023PhRvD.107l3027Z}
\begin{equation}
    Q_{\rm e}=\frac{\rho(r)}{m_{\chi}\tau}\frac{{\rm d}N_{\rm e}}{{\rm d}E},
\end{equation}
where $m_{\chi}$ and $\tau$ are the mass of the HDM particle and its decay lifetime. The electron spectrum per decay ${\rm d}N_{\rm e}/{\rm d}E$ is calculated by code {\tt HDMSpectra} \cite{2021JHEP...06..121B}.

For PBHs \cite{2021JCAP...03..011D},
\begin{equation}
    Q_{\rm e}=\frac{f_{\rm PBH}\rho(r)}{M_{\rm PBH}}\frac{{\rm d}N_{\rm e}}{{\rm d}E{\rm d}t},
\end{equation}
where $M_{\rm PBH}$ and $f_{\rm PBH}$ are the mass of memory-burdened PBHs and the fraction of which are composed of the total DM today. The electron spectrum from Hawking radiation ${\rm d}N_{\rm e}/({\rm d}E{\rm d}t)$ is calculated by code {\tt BlackHawk} \cite{2019EPJC...79..693A,2021EPJC...81..910A,2021JHEP...06..121B}. In addition, secondary emission of Hawking radiation is also considered in this work.

For $\gtrsim 10$ TeV electrons, the diffusion length is short enough that the diffusion could be neglected as demonstrated in \cite{2023PhRvD.107l3027Z}, indicating that the approximate solution of equation \ref{eq:prop} has the form of 
\begin{equation}
    F_{\rm DM}(E,r)=\frac{c}{4\pi}\frac{{\rm d}n_{\rm e}}{{\rm d}E}=\frac{c}{4\pi}\frac{1}{b(E)}\int_E dE'Q_{\rm e}(E',r).
\end{equation}
Furthermore, $\rho(r)=0.4\,\rm GeV/cm^3$ is also adopted as the local DM density since only the electrons near the Earth could be detected before cooling.

The energy-loss term $b(E)$ in the Milky Way could be approximated as \cite{1995PhRvD..52.3265A}
\begin{equation}
    b(E)=b_0+b_1(E/1\, {\rm GeV})+b_2(E/1 \,{\rm GeV})^2
\end{equation}
due to ionization ($b_0\approx 3\times10^{-16}\,\rm GeV/s$), bremsstrahlung ($b_1\approx 10^{-15}\,\rm GeV/s$), synchrotron radiation and inverse Compton process ($b_2\approx 10^{-16}\,\rm GeV/s$), respectively.

\begin{figure}
    \centering
    \includegraphics[width=1\linewidth]{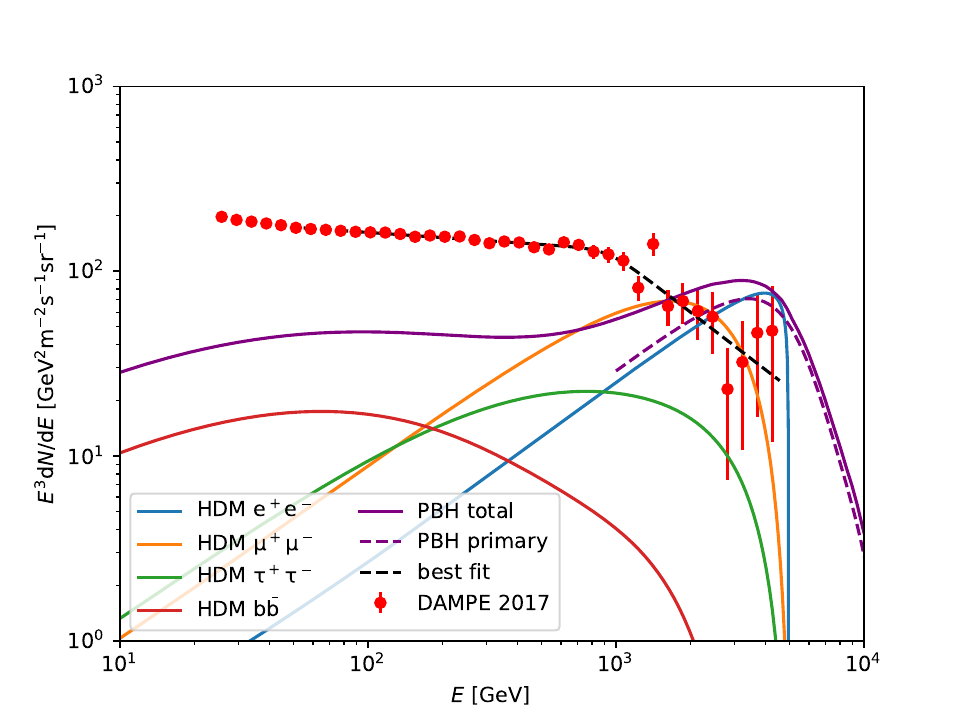}
    \caption{The DAMPE measurements (red points) and the expected spectra from HDM and memory-burdened PBHs. The blue, orange, green, and red lines represent the 10 TeV HDM decay spectra for different channels (decay lifetime $\tau=10^{26}$ s). The purple lines represent the total (solid) and primary (dashed) emissions of $10^{10}\,{\rm g}$ PBHs' Hawking radiation ($k=0.75$, $f_{\rm PBH}=1$).  The black dashed line represents our best smoothly broken power-law fit to the DAMPE measurements in the energy range from 55 GeV to 4.57 TeV.}
    \label{fig:spectra}
\end{figure}

The cosmic-ray electron measurements of DAMPE between 27 December 2015 and 8 June 2017 are adopted in this work, which is fitted in the energy range from 55 GeV to 4.57 TeV by a smoothly broken power-law (SBKPL) model \cite{2017Natur.552...63D},
\begin{equation}
    F_{\rm bkg}=\phi_0\left(\frac{E}{100 \,{\rm GeV}}\right)^{-\gamma_1}\left[ 1+\left(\frac{E}{E_b}\right)^{-\frac{\gamma_1-\gamma2}{\Delta}}\right]^{-\Delta}
\label{eq:sbkpl}
\end{equation}
with the smoothness parameter $\Delta$ fixed to be 0.1 \cite{2013Sci...339..807A}.
This SBKPL spectrum is treated as the background in our analysis. Previous studies have shown that the observational data around 1 TeV exceeds the GALPROP background, and there may exist a bump component near 1 TeV \cite{2021JCAP...05..012Z}. However, for the purposes of our work, even if this component originates from dark matter, it does not belong to the DM mass range we are considering and can still be treated as background.
We also plot the expected spectra from 10 TeV HDM (decay lifetime $\tau=10^{26}$ s) and $10^{10}$ g memory-burdened PBHs (entropy index $k = 0.75$) to be contrasted in Fig. \ref{fig:spectra}.

Then the total electron flux including the DM component is
\begin{equation}
    F_{\rm tot}=F_{\rm bkg}+F_{\rm DM}.
\end{equation}
The expected electron counts can be expressed as
\begin{equation}
    \mu_k=T_{\rm o}\int_k A F_{\rm tot}{\rm d}E
\end{equation}
with the energy bin $k$, the operation time $T_{\rm o}$ and the acceptance of DAMPE $A$, which are extracted from \cite{2017Natur.552...63D}.
The generalized $\chi^2$ has the form of
\begin{equation}
    \chi^2=\sum_{k}\frac{[n_k-\mu_k(\theta,\theta_{\rm bkg})]^2}{\sigma_k^2},
\end{equation}
where $n_k$ is the measured electron counts, $\sigma_k$ is the data error, and $\theta$ is the DM parameter for a given DM mass. 
The $\theta_{\rm bkg}$ represents the spectral parameters in Eq.~(\ref{eq:sbkpl}) except $\Delta$, which are free to vary in our analysis.
The $\theta$ could be the decay lifetime $\tau$ for HDM, or the entropy index $k$ / the fraction $f_{\rm PBH}$ for memory-burdened PBHs.
The constraints for HDM and memory-burdened PBHs at $95\%$ confidence level can be derived from $\Delta\chi^2=2.71$ compared with the best-fit results. The code {\tt iminuit} \cite{2022zndo...3949207D} is used for the fitting process.

It is natural for HDM to assume that all particles have the same mass, whereas PBHs are generally expected to have an extended mass function in an inflationary scenario. Although an exact monochromatic mass function is not physically realistic, it could be a suitable choice to approximately apply for various extended mass distributions \cite{2010PhRvD..81j4019C}. Nevertheless, the monochromatic mass function is adopted for the memory-burdened PBHs in this work.

\begin{figure}
    \centering
    \includegraphics[width=1\linewidth]{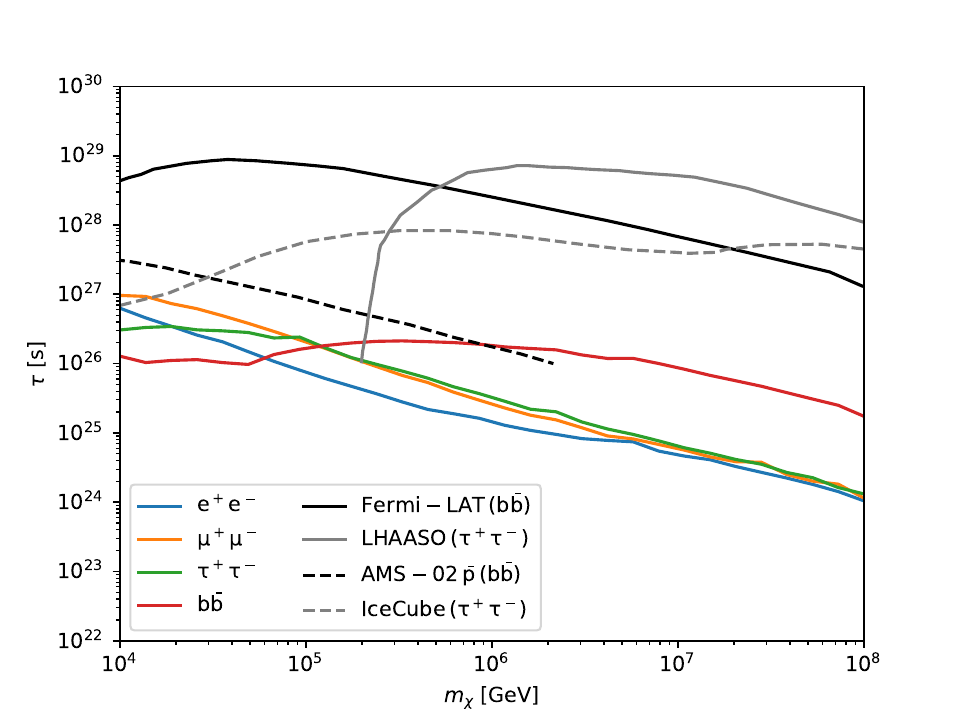}
    \caption{The lower limits on the HDM decay lifetime at $95\%$ confidence level for the $e^+e^-$ (blue), $\mu^+\mu^-$ (orange), $\tau^+\tau^-$ (green), and $b\bar{b}$ (red) channels. The black and gray solid lines represent the constraining results from the Fermi-LAT \cite{2020JCAP...01..003I} and LHAASO \cite{2022PhRvL.129z1103C} gamma-ray observations, respectively. The black and gray dashed line represent the results from the AMS-02 antiproton measurements \cite{2016PhRvL.117i1103A} and the IceCube neutrino measurements \cite{2018EPJC...78..831A}, respectively.}
    \label{fig:hdm_results}
\end{figure}

\section{Results}
\label{sec4}

The constraints on HDM decay lifetime are derived for the $e^+e^-$, $\mu^+\mu^-$, $\tau^+\tau^-$ and $b\bar{b}$ channels at $95\%$ confidence level in Fig. \ref{fig:hdm_results}, where the HDM particle mass ranges from 10 TeV to $10^5$ TeV. 
The most stringent results from the extragalactic photons measured by Fermi-LAT \cite{2015ApJ...799...86A} ($b\bar{b}$ channel), LHAASO \cite{2023PhRvL.131o1001C} ($\tau^+\tau^-$ channel) and IceCube \cite{2018EPJC...78..831A} ($\tau^+\tau^-$ channel) are plotted in Fig. \ref{fig:hdm_results} \cite{2020JCAP...01..003I,2022PhRvL.129z1103C}. Besides, we also draw the exclusion line from the AMS-02 cosmic-ray antiproton measurements ($b\bar{b}$ channel) to be contrasted \cite{2016PhRvL.117i1103A}.

We find that the results from $b\bar{b}$ channels are more stringent than those from other leptonic channels when HDM mass $\gtrsim$ 100 TeV. The main reason is the spectrum from $b\bar{b}$ channel is much softer than that from other channels. For the situation of higher HDM mass, the spectrum from $b\bar{b}$ channel would be more likely close to the DAMPE measurements while the spectra from other channels only left a tail.

\begin{figure}
    \centering
    \includegraphics[width=1\linewidth]{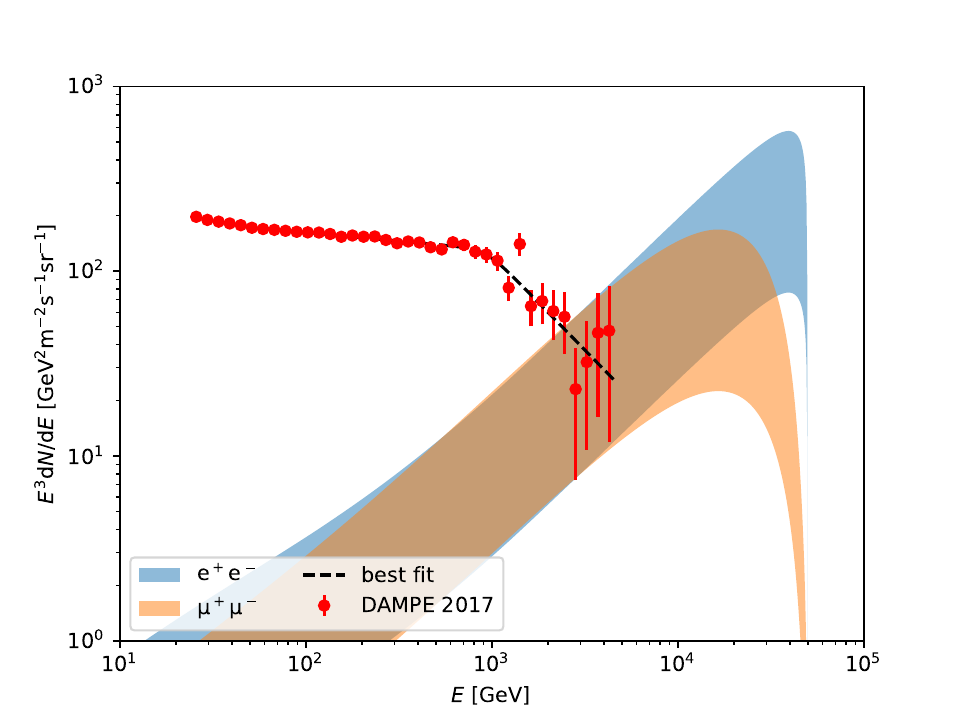}
    \caption{The high-energy tail ($\sim2.63-4.57\,{\rm TeV}$) of the DAMPE electron spectrum shows an upward trend, possibly suggesting the presence of an additional component. We discuss the possibility that this spectral behavior is real and is due to HDM or PBH. 
    This plot shows examples (100 TeV HDM decaying through $\rm e^+e^-$ or $\rm \mu^+\mu^-$ channels) of HDM spectra that can match the last 4 points of the DAMPE measurements (i.e., all spectra below and above the boundaries of the shaded region can be regarded as consistent with the data). The black dashed line represents our best smoothly broken power-law fit to the DAMPE measurements in the energy range from 55 GeV to 4.57 TeV.}
    \label{fig:example}
\end{figure}

\begin{figure}[t]
    \centering
    \includegraphics[width=1\linewidth]{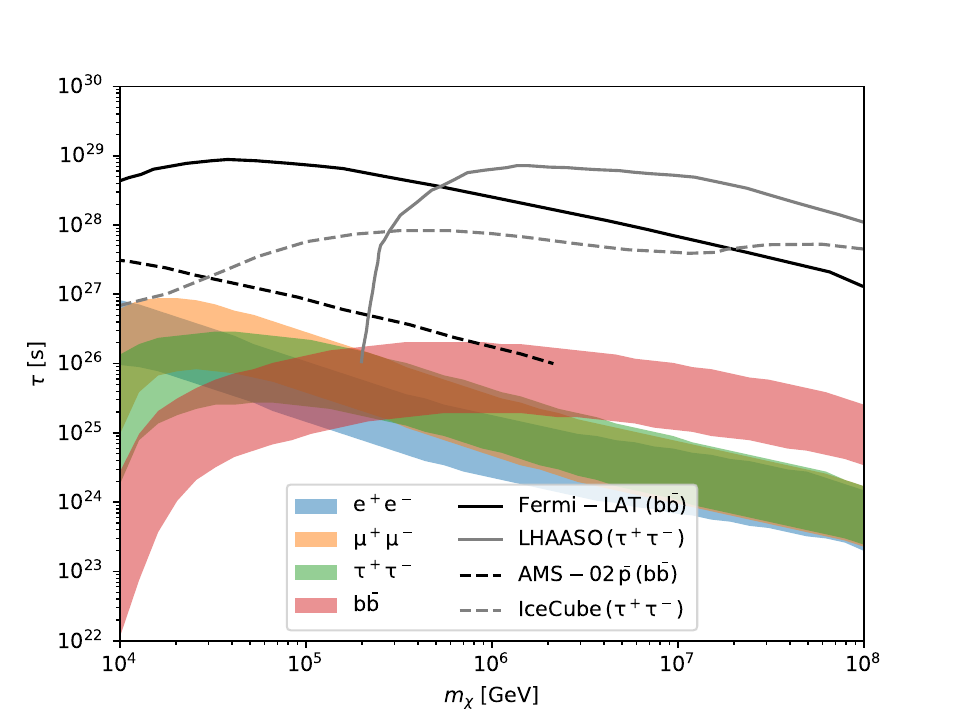}
    \caption{The required parameter space for using HDM decay to account for the spectral upturn at the high-energy tail of the DAMPE electron spectrum. The blue, orange, green, and red regions are for $e^+e^-$, $\mu^+\mu^-$, $\tau^+\tau^-$, and $b\bar{b}$ channels, respectively. The black solid line, gray solid line, black dashed line and gray dashed line represent the constraining results from Fermi-LAT \cite{2020JCAP...01..003I}, LHAASO \cite{2022PhRvL.129z1103C}, the AMS-02 antiproton \cite{2016PhRvL.117i1103A}, and the IceCube neutrino \cite{2018EPJC...78..831A}, respectively. We can see that the required parameters have been ruled out by existing constraints.}
    \label{fig:signals}
\end{figure}

We noticed that the DAMPE electron measurements of the energy range $\sim2.63-4.57\,{\rm TeV}$ show a sign of spectral upturn compared with those of the lower energy (though not statistically significant), which could possibly be contributed by an additional DM component.
We assume that it does come from the DM component and try to find out the DM parameter space of HDM and memory-burdened PBHs by simply requiring that the DM spectra should not exceed all four error bar ranges (i.e., the DM model spectrum needs to cross at least one error bar) within the energy range of $\sim2.63-4.57\,{\rm TeV}$.
Examples of DM spectra of 100 TeV HDM that meet the requirement are shown in Fig. \ref{fig:example}.
The required parameter space of HDM for different channels is plotted in Fig. \ref{fig:signals}.
It is obvious that the parameter space of the tentative DM component is totally excluded by existing observations.

Since PBH mass of $10^{10}-10^{13}$ g would be deeply constrained by BBN \cite{2024MNRAS.532..451T}, we only obtained the results of $< 10^{10}$ g PBHs.
The Hawking radiation would be stronger if the PBH mass is lower, thus the suppression from the memory burden effect must be more efficient to ensure the low-mass PBHs could still be composed of the entire DM. Therefore, the constraining results of $k$ would be very large for PBHs of $\lesssim 10^3$ g, which is adopted as the lower mass boundary.

\begin{figure}
    \centering
    \includegraphics[width=1\linewidth]{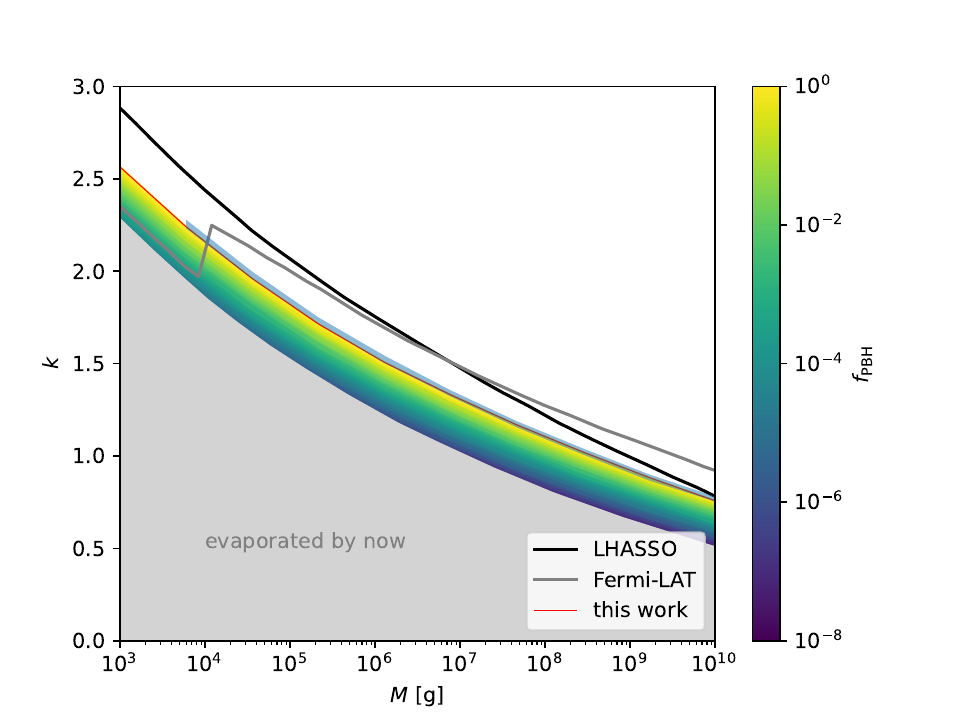}
    \caption{The colored region is the result derived in this work using the DAMPE electron measurements, with the color representing the upper limits on $f_{\rm PBH}$. The solid lines with legends show the lower limits of entropy index $k$ for memory-burdened PBHs at $95\%$ confidence level when the fraction $f_{\rm PBH}=1$. The black and gray lines represent the constraints derived from the Fermi-LAT and LHAASO measurements of the extragalactic gamma-ray background, respectively \cite{2024MNRAS.532..451T}.
    The very narrow blue region is the required parameter space for using PBHs' Hawking evaporation to account for the spectral upturn at the high-energy tail of the DAMPE electron spectrum.}
    \label{fig:pbh_results}
\end{figure}

The parameters of memory-burdened PBHs are constrained at $95\%$ confidence level in Fig.~\ref{fig:pbh_results}. The gray region means the memory-burdened PBHs are completely evaporated by now and do not survive to be a DM candidate. The colored region represents the upper limits on $f_{\rm PBH}$ derived in this work using the DAMPE electron measurements. 
For a given set of $M$ and $k$, the corresponding colorbar value indicates the upper limit of $f_{\rm PBH}$.
The red solid line in Fig.~\ref{fig:pbh_results} is the upper boundary of the color region and thus indicates the lower limits on the entropy index $k$ of the memory-burdened PBHs at $95\%$ confidence level when the fraction $f_{\rm PBH} = 1$. 
The most stringent results are from Fermi-LAT and LHAASO as well, which are constrained in \cite{2024MNRAS.532..451T}. 
Note that there is a kink around $M_0\sim10^4\,{\rm g}$ for the Fermi-LAT result, the reason is that the authors of Ref.~\cite{2024MNRAS.532..451T} do not trust the Hawking secondary emission at energies of $\lesssim10^{-6} ~kT$ and only adopt primary emission in this energy range (while above $10^{-6} ~kT$ using both primary+secondary). However, we have incorporated both primary and secondary Hawking emissions in our analysis even at energies of $\lesssim10^{-6}~kT$. If the same treatment is applied, our results would also exhibit a kink at approximately $\sim3-4\times10^3\,{\rm g}$.

Fig.~\ref{fig:k=2} is a slice of Fig.~\ref{fig:pbh_results} corresponding to the case of $k = 2$, for better illustration of the constraints on $f_{\rm PBH}$.
In both figures, the required parameter spaces of the memory-burdened PBHs able to account for the spectral upturn at the high-energy tail of the DAMPE spectrum are also shown as the blue regions, which are also totally excluded by existing results.

\section{Discussion and summary}
\label{sec5}

We've just discussed the case where HDM or memory-burdened PBHs are the sole dark matter candidates in this work. However, the possibility remains that both could be part of the dark sector. 
PBHs are believed to be capable of emitting DM particles, due to the fact that Hawking radiation can emit any particle whose Compton wavelength is comparable to the Schwarzschild radius \cite{1975CMaPh..43..199H}. 
Since for traditional PBHs only $\gtrsim 5 \times 10^{14}\,{\rm g}$ black holes can survive from the evaporation to the present day, whose temperature is $k_{B}T \lesssim 100\,{\rm MeV}$, the emitted DM particles are generally supposed to be the fuzzy DM \cite{2024arXiv240402956D,2021PhRvD.104g5007B,2024EPJC...84..723L}. 
For memory-burdened PBHs, however, one might expect Hawking radiation to emit HDM particles because the black holes can have very small masses, and are therefore capable of producing very massive particles. In that case, the annihilation signals would be enhanced by the DM profile surrounding the PBHs, while the decay signals would be independent of the profile.

One should also note that the memory burden effect is based on the assumption that Hawking radiation is purely thermal. However, if considering the PBH background as dynamical and the energy conservation, Hawking radiation could be treated as a semi-classical tunneling process, indicating that the radiation would not be exactly thermal \cite{2000PhRvL..85.5042P,2006PhRvD..73f4003J,2006tmgm.meet.1585P}.
Therefore, we believe that the constraining results (both ours and the ones in \cite{2024MNRAS.532..451T,2024arXiv241007604C}) at the mass window of $\lesssim 10^{10}\,{\rm g}$ are quite model-dependent.

\begin{figure}
    \centering
    \includegraphics[width=1\linewidth]{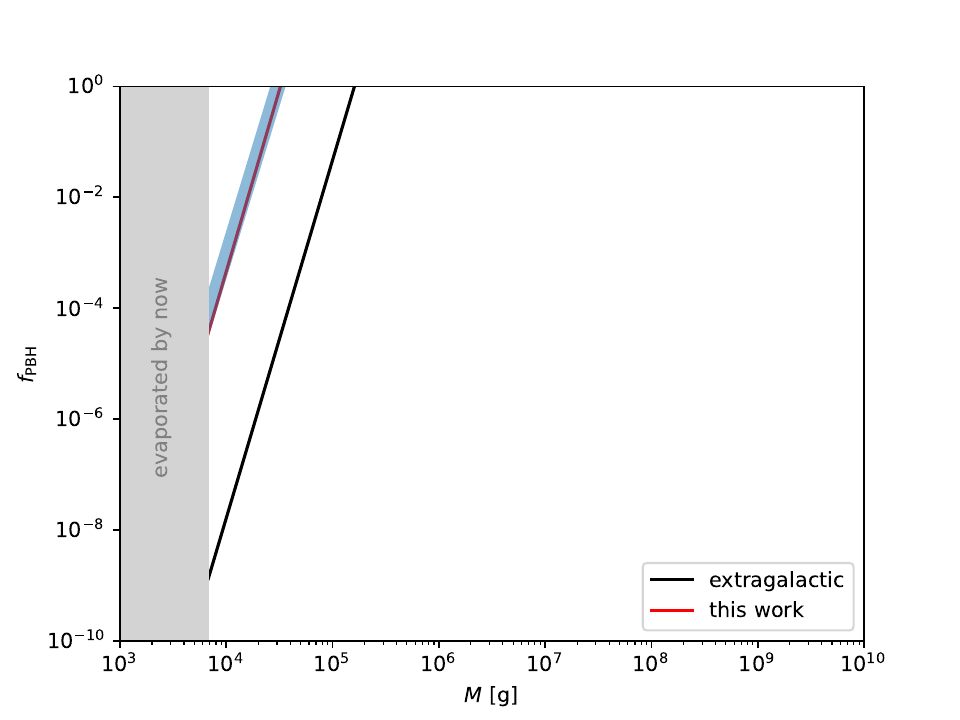}
    \caption{The upper limits on the $f_{\rm PBH}$ for memory-burdened PBHs at $95\%$ confidence level when k=2 (red line), where $f_{\rm PBH}$ is the fraction of dark matter in the form of PBHs. The black line represents the upper limits derived from the Fermi-LAT and LHAASO measurements of the diffuse extragalactic gamma-ray background \cite{2024MNRAS.532..451T}. 
    The blue region is the required parameter space for using PBHs' Hawking evaporation to account for the spectral upturn at the high-energy tail of the DAMPE electron spectrum.}
    \label{fig:k=2}
\end{figure}

\begin{figure}
    \centering
    \includegraphics[width=1\linewidth]{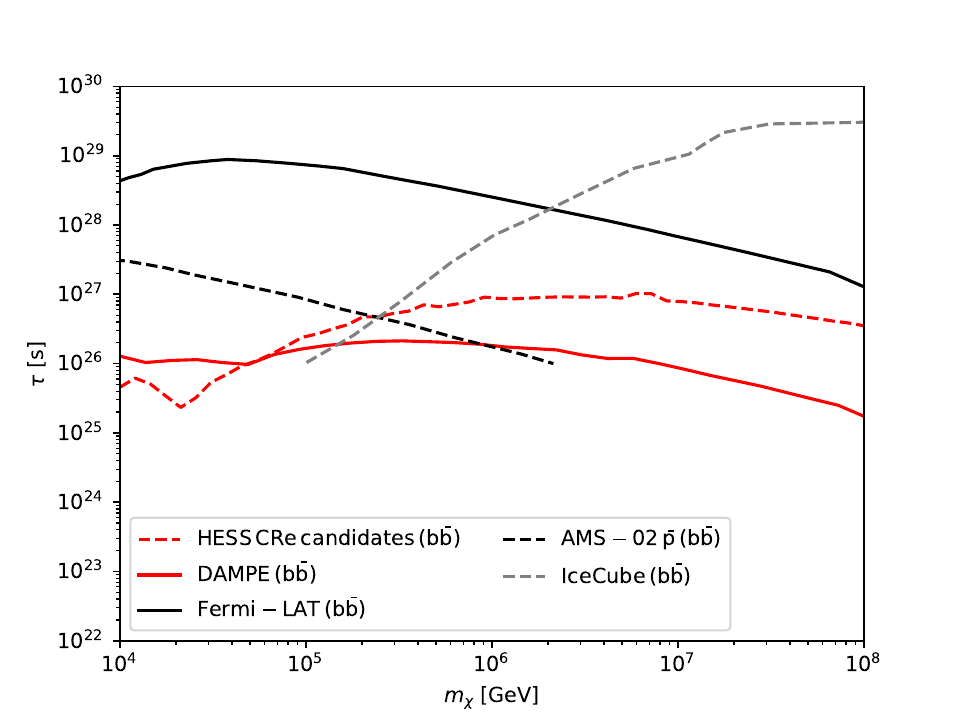}
    \caption{ Tentative lower limits on the HDM decay lifetime for a $b\bar{b}$ channel derived from the spectrum of HESS electron candidate events \cite{2024PhRvL.133v1001A}. 
    Note that this result should not be treated as definitive limits—instead, it serves as an assessment of potential future constraints that could be achieved. See the main text for details.
    The results through the $b\bar{b}$ channel derived from Fermi-LAT \cite{2020JCAP...01..003I}, AMS-02 \cite{2016PhRvL.117i1103A}, and IceCube \cite{2018EPJC...78..831A} are also plotted for comparisons.} 
    \label{fig:hess}
\end{figure}

Recently, the HESS collaboration reported their measurements of the spectrum of {\it cosmic-ray electron candidate events} \cite{2024PhRvL.133v1001A}. 
However, since the results of HESS were derived by an indirect method and there is some controversy regarding the high-energy part of their results (even for a spectrum of candidates), we have not taken them into consideration in this paper. If the HESS results were reliable, they would provide stronger constraints on HDM and PBH.
In Fig.~\ref{fig:hess}, we tentatively present constraints derived from HESS's energy spectrum, but it should be noted that these results should not be interpreted as definitive limits—instead, they serve as an assessment of potential future constraints that could be achievable. As can be found, in such a case, the constraints derived from electron observations remain weaker than those obtained through gamma-ray and neutrino observations.

{\it Summary.} In this paper, we use the electron measurements from DAMPE to constrain the parameters of HDM and memory-burdened PBHs. 
We first calculate the expected electron flux from DM (see Fig. \ref{fig:spectra}) by solving the propagation equation. Since the electron energy is very high, we adopt an approximate solution by ignoring the diffusion and a constant DM density near the Earth.
Then the $\chi^2$ is calculated by fitting the model flux to the DAMPE data to obtain the lower limits on the HDM decay lifetime at $95\%$ confidence level (Fig. \ref{fig:hdm_results}). 
The lower limits on the entropy index of memory-burdened PBHs are also derived by a similar procedure (Fig. \ref{fig:pbh_results}). We also derive the constraints on the fraction $f_{\rm PBH}$ when $k=2$ (Fig. \ref{fig:k=2}).
Although our final results are weaker than the constraints based on the gamma-ray and neutrino measurements from Fermi-LAT, LHAASO and IceCube, however are stronger than the ones from the AMS-02 antiproton measurements at $m_\chi>10^6\,{\rm GeV}$. As another messenger in addition to electromagnetic waves and neutrinos, the cosmic-ray electron measurements provide another way to robust the existing constraints. 
In the future, the upcoming updated electron measurements of DAMPE will strengthen the results of this work. Furthermore, the electron measurements of VLAST, the next-generation satellite mission to detect gamma rays and cosmic rays proposed by the DAMPE group people \cite{2022AcASn..63...27F}, may also provide more stringent constraints. 

In addition, we note that the high-energy tail ($\sim2.63-4.57\,{\rm TeV}$) of the DAMPE electron spectrum shows a sign of spectral upturn (statistically insignificant), possibly suggesting the presence of an additional component (see Fig. \ref{fig:signals}). We consider the possibility that this spectral behavior is due to HDM or PBH and try to derive what parameter space can accommodate the observational data in the $\sim2.63-4.57\,{\rm TeV}$ under the assumption that the upturn structure is real. We find that the required parameters have been completely ruled out by the existing constraints (Fig. \ref{fig:signals} - \ref{fig:k=2}), suggesting that this structure, even if it were real, could not be attributed to HDM decay or PBHs.

\section{Acknowledgement}

We greatly acknowledge the helpful discussions with Xiang Li, Sen Guo, Ming-Xuan Lu, Xing-Fu Zhang, and Yu Wang. This work is supported by the National Key Research and Development Program of China (Grant No. 2022YFF0503304), and the Guangxi Talent Program ("Highland of Innovation Talents").

\bibliographystyle{apsrev4-1-lyf} 
\bibliography{refs}

\end{document}